\documentclass[a4paper,11pt]{article}
\usepackage{jheppub}
\usepackage{lineno}
\usepackage{braket}
\usepackage{amsmath}
\usepackage{amssymb}
\usepackage{url}
\usepackage{graphicx}
\usepackage{subcaption}
\usepackage{booktabs}
\usepackage{xcolor}
\graphicspath{{figs/}{../}}
\usepackage[section]{placeins}

\newcommand{\eq}[1]{Eq.~(\ref{#1})}
\newcommand{\eqs}[2]{Eqs.~(\ref{#1})--(\ref{#2})}

\newcommand{\sigm}{\sigma_{\nu\chi}/m_\chi}
\newcommand{\Rem}{R_{\rm em}}
\newcommand{\Rsp}{R_{\rm sp}}

\title{\boldmath Probing flavor-dependent neutrino–dark matter interactions with the flavor composition of high-energy cosmic neutrinos}

\author[a]{Ki-Young Choi,}
\author[a]{Ankhbayar Tugssaikhan}

\affiliation[a]{Department of Physics and Institute of Basic Science, Sungkyunkwan University, Suwon 16419, Korea}

\emailAdd{kiyoungchoi@skku.edu}
\emailAdd{gunj0826@g.skku.edu}

\abstract{
High-energy astrophysical neutrinos probe non-standard neutrino interactions over cosmological distances. We study flavor-dependent interactions between neutrinos and dark matter (DM) for neutrinos produced inside the dense DM spike surrounding the supermassive black hole of the blazar TXS~0506+056. A flavor-diagonal potential $V_\alpha=A_\alpha\,\rho_\chi(r)$ induced by coherent forward scattering modifies the mixing in the medium, and the flavor-dependent scattering cross section attenuates the neutrino flux along the line of sight. We show that the propagation is adiabatic throughout the spike and the halo, so that the matter eigenstates produced at the source are conserved and only re-projected onto flavors at Earth. Two regimes emerge. For $A\ll 10^{-38}\,{\rm cm^3}$ the mixing is unaffected and the IceCube three-flavor composition limits the cross section to $\sigm\lesssim 10^{-28}\,{\rm cm^2/GeV}$. For larger $A$ the interacting flavor is isolated into a single matter eigenstate at the source; the two remaining eigenstates escape without attenuation, producing a survival plateau that relaxes the limits by up to four orders of magnitude. The sign of the potential can differ between neutrinos and antineutrinos, so that the isolated state connects to different vacuum mass eigenstates for $\nu$ and $\bar\nu$, which leaves a characteristic imprint on the flavor composition even without attenuation. The requirement that the potential induced by the dark matter density around Earth do not affect the observed PMNS matrix, so that the flavor projection at Earth is performed with the local matter mixing, gives A$\lesssim 10^{-30}\text{cm} ^3$. The combined constraints are presented in the $(A,\sigm)$ plane and compared with a vector-mediator benchmark. 
}

\begin{document}
\maketitle
\flushbottom

\section{Introduction}
\label{Intro}

The existence of the neutrino has been accompanied by mystery from its proposal in the 1930s up until now. Wolfgang Pauli first proposed the existence of the neutrino to explain the continuous beta-decay spectrum and preserve the conservation of energy and angular momentum~\cite{Brown:1978idea}. In 1957, Wu's experiment showed that the weak interaction violates parity~\cite{Wu:1957my}, to which  John Blatt reacted by writing to Wolfgang Pauli? ``$[\cdots]$ We are all rather shaken by the death of our well-beloved friend, parity''~\cite{vonMeyenn19792005}. A defining feature of modern neutrino physics is the mismatch between the weak-interaction flavor eigenstates and the propagation mass eigenstates, parameterized by the unitary Pontecorvo--Maki--Nakagawa--Sakata (PMNS) matrix~\cite{Maki:1962mu,Pontecorvo:1967fh}. Neutrinos, though very light, experience flavor oscillations~\cite{SNO:2001kpb}, and when propagating through background matter they experience coherent forward scattering that alters their dispersion relation and flavor evolution, the Mikheyev--Smirnov--Wolfenstein (MSW) effect~\cite{Wolfenstein:1977ue,Mikheyev:1985zog}.

Converging lines of evidence indicate that dark matter (DM) constitutes the dominant share of the matter budget of the Universe and that galaxies reside inside DM halos~\cite{Clowe:2006eq,Rubin:1980zd}. Should neutrinos couple to DM, two distinct effects follow. Coherent forward scattering off the DM background produces an MSW potential that modifies the mixing in the medium~\cite{Capozzi:2018bps,Choi:2019zxy,Choi:2020ydp,Chun:2022qcg}.
Incoherent scattering removes neutrinos from the beam, and has been used to constrain the neutrino–DM cross section with astrophysical sources~\cite{Reynoso:2016hjr,Arguelles:2017atb,Kelly:2018kma,Choi:2019ixb,Murase:2019xqj,Ferrer:2022xlw,Cline:2022qld}. The attenuation limits assume flavor-blind couplings. Flavor-dependent medium effects from coherent forward scattering have only recently been considered~\cite{Tseng:2026evn,Dev:2026kqu, Abbaslu:2026row}. Here we treat both effects together for a flavor-dependent coupling: the change of mixing inside the medium as well as the flavor-dependent suppression of the flux. 
The interaction that attenuates the flux also fixes which flavor combinations propagate as eigenstates in the dense DM environment, and this changes the attenuation qualitatively.

Thanks to IceCube~\cite{IceCube2012,PhysRevLett.113.101101,IceCube:2013low} and KM3NeT~\cite{Coniglione_2015}, high-energy astrophysical neutrinos are now detected routinely. IceCube has recently characterized the flavor composition of the all-sky flux between $5$~TeV and $10$~PeV, with best fit $(f_e:f_\mu:f_\tau)_\oplus\approx(0.30:0.37:0.33)$~\cite{IceCube:2025xyz}. Standard astrophysical scenarios assume production through charged-pion decay~\cite{BECKER2008173}, with a source composition $(1:2:0)_S$ that oscillates to $\approx(0.33:0.34:0.33)_\oplus$ with the oscillation parameters of~\cite{Esteban:2020cvm}. In sources with strong magnetic fields the muons cool before decaying and the composition becomes muon damped, $(0:1:0)_S$~\cite{Kashti:2005qa} that oscillates to $\approx(0.22:0.41:0.38)_\oplus$. The origin of the diffuse neutrino flux remains uncertain~\cite{IceCube:2019vua,IceCube:2020acn}. The blazar TXS~0506+056~\cite{Padovani_2019} and the Seyfert galaxy NGC~1068 are the best-established candidate sources. We adopt TXS~0506+056, whose supermassive black hole is expected to be surrounded by a dense DM spike, as our benchmark, following~\cite{Ferrer:2022xlw}.

We find that the propagation is adiabatic throughout the spike and the Galactic halo, so the matter eigenstates produced at the source are conserved and only re-projected onto flavors at Earth. For a large enough potential, the interacting flavor is isolated into a single eigenstate that is attenuated, while the other two escape. This relaxes the limits on the cross section by about three orders of magnitude compared with the flavor-blind case. The potential alone can already shift the flavor composition, differently for neutrinos and antineutrinos.

The paper is organized as follows. Section~\ref{sec:framework} sets up the DM profile, the effective potential, the adiabatic propagation including the Milky Way halo, and the sign difference between neutrinos and antineutrinos. Section~\ref{sec:flavor} presents the flavor composition at Earth induced by the potential alone. Section~\ref{sec:attenuation} introduces flavor-dependent attenuation and derives the limits on $\sigm$ as a function of the potential strength and combines both in the $(A,\sigm)$ plane and compares with a mediator benchmark. We conclude in Section~\ref{sec:conclusion}.

\section{Framework}
\label{sec:framework}

\subsection{Dark matter profile around TXS 0506+056}

\begin{figure}[!t]
    \centering
    \includegraphics[width=1\linewidth]{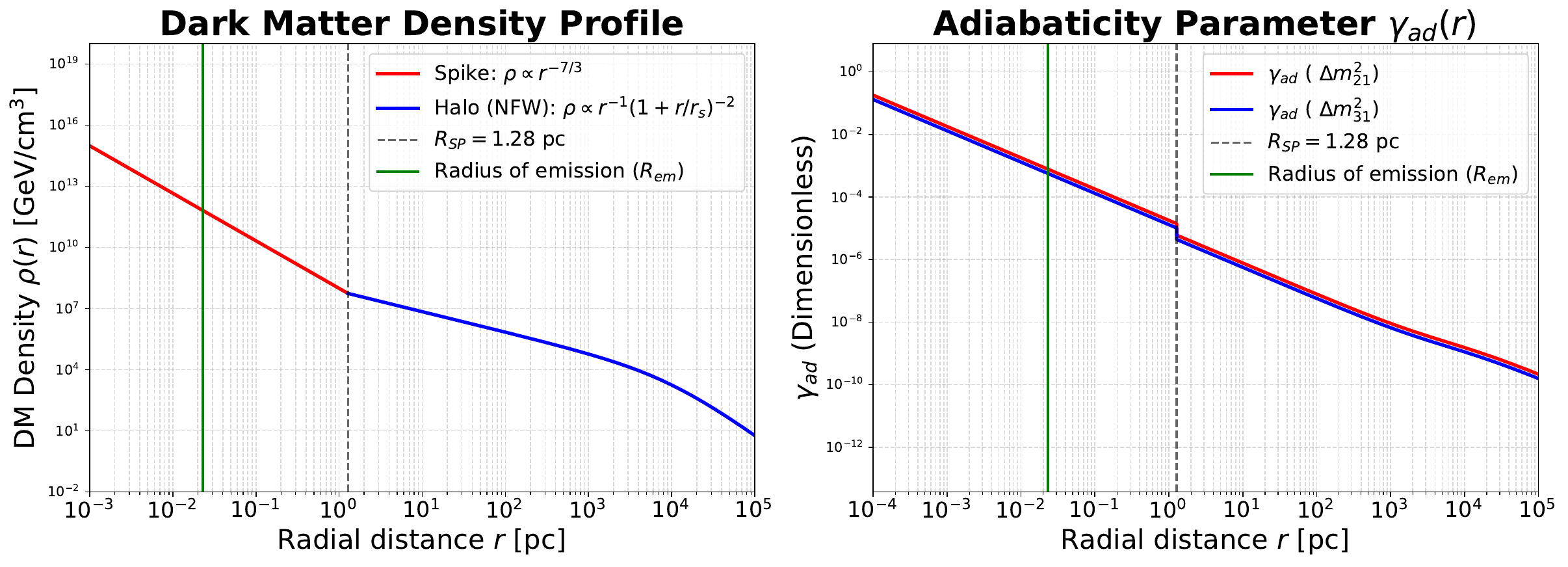}
    \caption{Left: DM density around the black hole of TXS~0506+056, \eq{eq:profile}. Right: the adiabaticity parameter $\gamma_{\rm ad}(r)$ of \eq{eq:adiabaticity} evaluated for $E=100$~TeV. The vertical lines mark the emission radius $\Rem=0.023$~pc and the spike radius $\Rsp=1.28$~pc. For a power-law profile $\gamma_{\rm ad}$ does not depend on the normalization of the potential, so the curve gives the adiabaticity at whichever radius the resonance falls for a given $A$; $\gamma_{\rm ad}\ll1$ everywhere outside $\Rem$.}
    \label{fig:adiabaticity}
\end{figure}

Dark matter concentrates into dense spikes around black holes that grow adiabatically~\cite{Quinlan:1994ed}. Following~\cite{Ferrer:2022xlw}, the density around the supermassive black hole of TXS~0506+056 is
\begin{equation}
    \rho_\chi(r)=
    \begin{cases}
        0 & r \leq 4R_S, \\
        \dfrac{\rho_{\rm sp}(r)\,\rho_{\rm sat}}{\rho_{\rm sp}(r)+\rho_{\rm sat}} & 4R_S < r < \Rsp, \\[6pt]
        \rho_0 \left(\dfrac{r}{r_0}\right)^{-\gamma}\left(1+\dfrac{r}{r_0}\right)^{-3+\gamma} & r \geq \Rsp,
    \end{cases}
    \label{eq:profile}
\end{equation}
with the spike $\rho_{\rm sp}(r)=\rho_R\,g_\gamma(r)\,(\Rsp/r)^{\gamma_{\rm sp}}$, $\gamma_{\rm sp}=(9-2\gamma)/(4-\gamma)=7/3$ for $\gamma=1$, $g_\gamma\simeq1$, and $\rho_R=\rho_0(\Rsp/r_0)^{-\gamma}$ ensuring continuity at the spike radius $\Rsp=\alpha_\gamma r_0\,[M_{\rm BH}/(\rho_0 r_0^3)]^{1/(3-\gamma)}$ with $\alpha_\gamma=0.1$. We take $M_{\rm BH}=3\times10^8M_\odot$, $\rho_0=7\times10^3\,{\rm GeV/cm^3}$, $r_0=10$~kpc, which give $\Rsp=1.28$~pc and a Schwarzschild radius $R_S=2.9\times10^{-5}$~pc. The saturation density $\rho_{\rm sat}=m_\chi/(\langle\sigma v\rangle t_{\rm BH})$ accounts for DM annihilation and is irrelevant for the non-annihilating DM considered here ($\rho_{\rm sat}\to\infty$). Neutrinos are emitted at $\Rem=0.023$~pc, where $\rho_\chi(\Rem)=6.4\times10^{11}\,{\rm GeV/cm^3}$. The column density from $\Rem$ to infinity is dominated by the spike,
\begin{equation}
    X_\chi\equiv\int_{\Rem}^{\infty}\rho_\chi(r)\,dr = 3.6\times10^{28}\,{\rm GeV/cm^2},
    \label{eq:column}
\end{equation}
which scales as $\Rem^{-4/3}$. The left panel of Fig.~\ref{fig:adiabaticity} shows the profile. For comparison, the Milky Way (MW) halo adds a column of only $\sim10^{22}\,{\rm GeV/cm^2}$, negligible for attenuation. 

\subsection{Effective potential and mixing in the medium}

In the flavor basis the phase difference hamiltonian of a neutrino of energy $E$ in the DM background is
\begin{align}
    \mathcal{H}&=\frac{1}{2E}\,U\mathcal{M}^2U^\dagger + V(r)
    = \frac{1}{2E}\,U_m\mathcal{M}_m^2U_m^\dagger ,
    \label{eq:H}\\
    V(r)&={\rm diag}\big(V_e(r),V_\mu(r),V_\tau(r)\big),\qquad V_\alpha(r)=A_\alpha\,\rho_\chi(r),
    \label{eq:V}
\end{align}
where $U$ is the vacuum PMNS matrix, $\mathcal{M}^2={\rm diag}(0,\Delta m_{21}^2,\Delta m_{31}^2)$, and $U_m$, $\mathcal{M}_m^2$ are the mixing matrix and effective mass-squared difference eigenvalues in the medium. The coefficient $A_\alpha$ (units of ${\rm cm^3}$ for $V$ in eV and $\rho_\chi$ in ${\rm eV/cm^3}$) encodes the forward-scattering amplitude of $\nu_\alpha$ on DM. We consider three scenarios in which only one flavor couples, $V_e$ only, $V_\mu$ only and $V_\tau$ only.\footnote{Only differences of the diagonal entries affect the oscillations, so a flavor-universal part of $V$ is unobservable in the flavor composition; it still contributes to the attenuation.}

For antineutrinos, the vacuum term is obtained with $U\to U^*$ and the sign of the potential depends on the interaction, medium, and energy of the neutrino~\cite{Ge:2018uhz,Choi:2019zxy,Choi:2020ydp}. Here, we simply consider two cases. First, neutrinos and anti neutrinos have potentials of opposite signs, $V\to-V$ for $\nu$ and $\bar\nu$. Second, $\nu$ and $\bar\nu$ have the same potentials of same sign. That is,
\begin{equation}
    \bar{\mathcal{H}}=\frac{1}{2E}\,U^*\mathcal{M}^2U^{T} \mp V(r)
    =\frac{1}{2E}\,\bar U_m\bar{\mathcal{M}}_m^2\bar U_m^\dagger ,
    \qquad \text{(opposite / same sign potential)},
    \label{eq:Hbar}
\end{equation}

where $\bar U_m$ and $\bar{\mathcal M}^2_m$ are the matter mixing matrix and effective mass-squared eigenvalues for antineutrinos, which coincide with $U_m^*$ and $\mathcal M^2_m$ only when $V\to0$. The sign of the potential is physically important, because it can change the flavor composition observed in the IceCube.

In this work, for the vacuum Hamiltonian, we use the global-fit values of the oscillation parameters (NuFIT~6.0, normal ordering)~\cite{Esteban:2024eli},
\begin{equation}
    \begin{aligned}
    \Delta m^2_{21}&=7.49\times10^{-5}\,{\rm eV^2}, &\qquad \Delta m^2_{31}&=2.51\times10^{-3}\,{\rm eV^2},\\
    \sin^2\theta_{12}&=0.308, &\qquad \sin^2\theta_{23}&=0.470, \qquad \sin^2\theta_{13}=0.022,
    \end{aligned}
    \label{eq:oscpar}
\end{equation}
with $\delta_{\rm CP}=212^\circ$. For a neutrino of energy $E$ the two vacuum frequencies are then
\begin{equation}
    \frac{\Delta m^2_{21}}{2E}=3.7\times10^{-19}\,{\rm eV}\left(\frac{100\,{\rm TeV}}{E}\right),
    \qquad
    \frac{\Delta m^2_{31}}{2E}=1.3\times10^{-17}\,{\rm eV}\left(\frac{100\,{\rm TeV}}{E}\right).
    \label{eq:vacfreq}
\end{equation}

It is instructive to compare the magnitude of the potential with the vacuum oscillation term to see the importance of the medium effect. 
At the radius of the neutrino source in the DM spike, where $\rho_\chi(\Rem)=6.4\times10^{11}\,{\rm GeV/cm^3}$, the potential is
\begin{equation}
    V(\Rem)=A\,\rho_\chi(\Rem)=6.4\times10^{-7}\,{\rm eV}\left(\frac{A}{10^{-27}\,{\rm cm^3}}\right),
    \label{eq:Vsource}
\end{equation}
where $A$ parametrize the magnitude of the potential.
Comparing with \eq{eq:vacfreq}, the mixing at the source is unaffected for $A\lesssim6\times10^{-40}\,{\rm cm^3}\left(\frac{100\,{\rm TeV}}{E}\right)$, the 1--2 sector is modified above this value, and both sectors are saturated when \ $2EV\gg\Delta m^2_{31}$, i.e. for $A\gtrsim2\times10^{-38}\,{\rm cm^3} \left(\frac{100\,{\rm TeV}}{E}\right)$.

In the detector on Earth, the local density of DM with $\rho_\oplus=0.4\,{\rm GeV/cm^3}$ gives
\begin{equation}
    V(\rho_\oplus)=A\,\rho_\oplus=4\times10^{-19}\,{\rm eV}\left(\frac{A}{10^{-27}\,{\rm cm^3}}\right),
    \label{eq:Vearth}
\end{equation}
which is of the same order as $\Delta m^2_{21}/2E$ for $A\sim10^{-27}\,{\rm cm^3}$ at 100~TeV. Requiring that the neutrinos oscillation experiment are consistent with the vacuum potential, 

we impose $A$ is much smaller not to affect the vacuum oscillation as
\begin{equation}
    A \ll \frac{\Delta m^2_{21}\cos2\theta_{12}}{2E\,\rho_\oplus}\simeq 4\times10^{-28}\,{\rm cm^3}\left(\frac{100\,{\rm TeV}}{E}\right).
    \label{eq:alpha_earth}
\end{equation}
Therefore, we keep $A$ up to $10^{-30}\,{\rm cm^3}$ in the following analysis.

\subsection{Adiabatic propagation}

If the DM density changes slowly along the path, each mass eigenstate tracks the local eigenvector of $\mathcal H$ without level crossing~\cite{Landau:1932ab,Zener:1932zz,Parke:1986jy}. The adiabaticity parameter at a resonance~\cite{Parke:1986jy,Dighe:1999bi,Lunardini:2000fy},
\begin{equation}
    \gamma_{\rm ad} \equiv \frac{2E\cos2\theta}{\Delta m^2\sin^22\theta}\,\frac{1}{\rho_{\rm res}}\left|\frac{d\rho_\chi}{dr}\right|_{\rm res} \ll 1,
    \label{eq:adiabaticity}
\end{equation}
is shown in the right panel of Fig.~\ref{fig:adiabaticity}. For a power law, $\rho^{-1}|d\rho/dr|=\gamma_{\rm sp}/r$ is independent of $\rho_0$ and of $A$. It is below $10^{-3}$ for $r\geq\Rem$ at 100~TeV. We therefore treat the whole path as adiabatic.

Let $f^S_\alpha$ be the flux ratio of the neutrino flavor $\alpha$ at the source. At the radius $\Rem$ it is projected onto the flux in the mass eigenstates,
\begin{equation}
    F_i = \sum_\alpha \big|(U_m^{S})_{\alpha i}\big|^2 f^S_\alpha ,
    \label{eq:Fi}
\end{equation}
where $U_m^S=U_m(\rho_\chi(\Rem))$ and the index $i$ labels the eigenvalues of $\mathcal H$ in increasing order. Adiabatic evolution preserves this ordering, so the $i$-th eigenstate at the source arrives as the $i$-th eigenstate at Earth. The flavor flux at Earth is then
\begin{equation}
    f^\oplus_\beta = \sum_i \big|(U_m^{\oplus})_{\beta i}\big|^2\,F_i ,
    \qquad U_m^\oplus=U_m(\rho_\oplus),
    \label{eq:fearth}
\end{equation}
which reduces to the familiar $f^\oplus_\beta=\sum_i|U_{\beta i}|^2F_i$ when \eq{eq:alpha_earth} holds.

Antineutrinos are treated in the same way with the Hamiltonian $\bar{\mathcal H}$ and the mixing matrix in the medium $\bar U_m$ of \eq{eq:Hbar}. For an antineutrino flavor flux $\bar f^S_\alpha$ at the source,
\begin{equation}
    \bar F_i = \sum_\alpha \big|(\bar U_m^{S})_{\alpha i}\big|^2 \bar f^S_\alpha ,
    \qquad \bar U_m^S=\bar U_m(\rho_\chi(\Rem)),
    \label{eq:Fibar}
\end{equation}
and at Earth
\begin{equation}
    \bar f^\oplus_\beta = \sum_i \big|(\bar U_m^{\oplus})_{\beta i}\big|^2\,\bar F_i ,
    \qquad \bar U_m^\oplus=\bar U_m(\rho_\oplus).
    \label{eq:fearthbar}
\end{equation}
In vacuum $|(\bar U)_{\alpha i}|^2=|U_{\alpha i}|^2$, so \eqs{eq:Fibar}{eq:fearthbar} coincide with \eqs{eq:Fi}{eq:fearth} for $A\to0$. However, in the case of opposite sign potential, the same flavor can be isolated in a different eigenstate (lightest instead of heaviest) and the eigenvalue ordering, which is what adiabatic evolution preserves, connects it to a different vacuum state. IceCube does not distinguish $\nu$ from $\bar\nu$, and the observable is the sum of the both
\begin{equation}
    \Phi^\oplus_\beta = f^\oplus_\beta + \bar f^\oplus_\beta ,
    \qquad
    f_\beta^{\rm obs}=\frac{\Phi^\oplus_\beta}{\sum_\gamma\Phi^\oplus_\gamma}.
    \label{eq:sum}
\end{equation}

\section{Flavor composition from the potential alone}
\label{sec:flavor}

\begin{figure}[!t]
    \centering
    \includegraphics[width=1\linewidth]{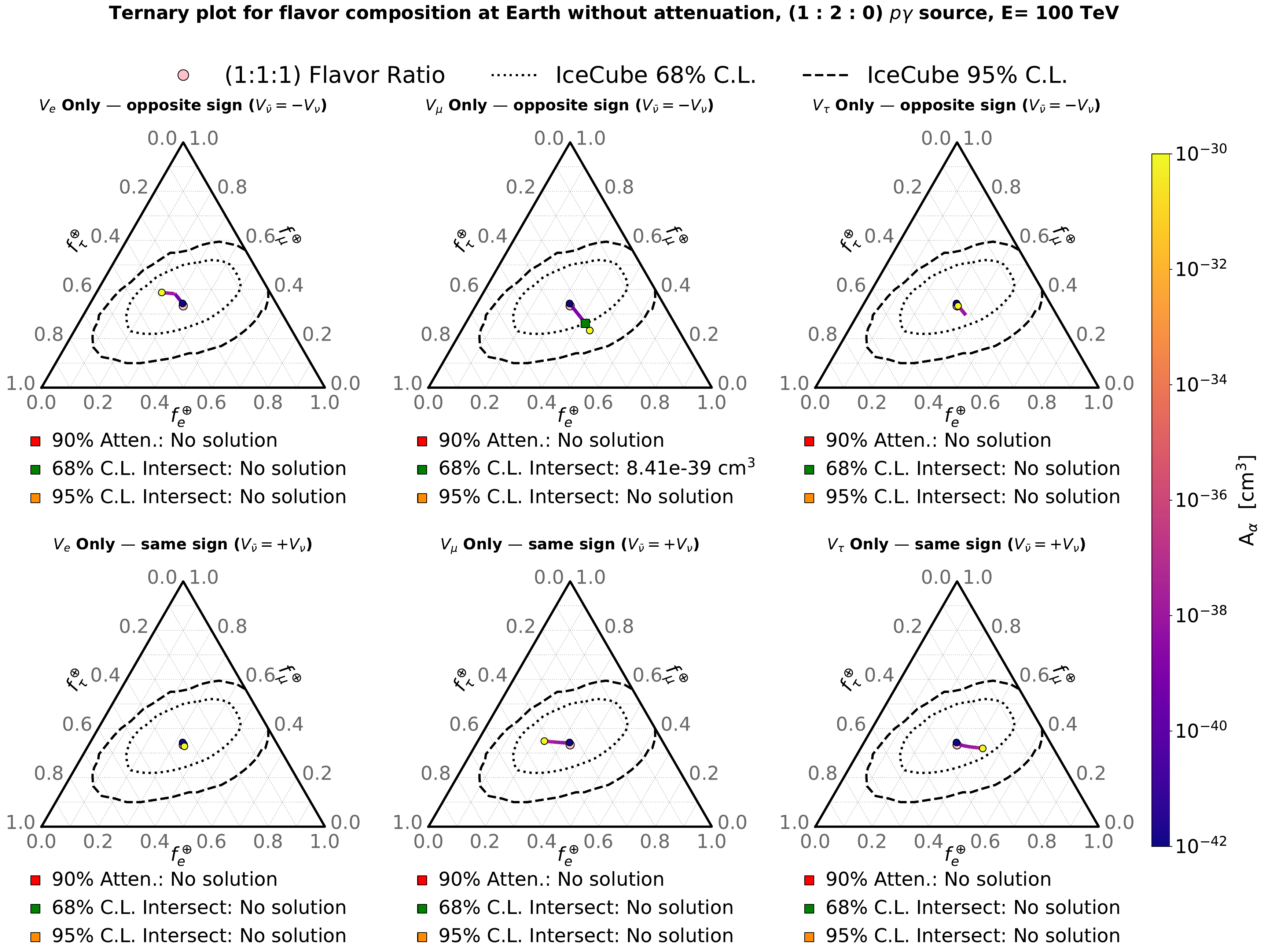}
    \caption{Ternary plot for flavor fractions at Earth, without attenuation, for a $(1:2:0)$ $p\gamma$ source at $E=100$~TeV (NO) as a function of the potential coefficient $A$ for $V_e$, $V_\mu$ and $V_\tau$ only (columns) and for potential of opposite sign and same sign (rows).}
    \label{fig:120_no_attenuation}
\end{figure}

\begin{figure}[!t]
    \centering
    \includegraphics[width=1\linewidth]{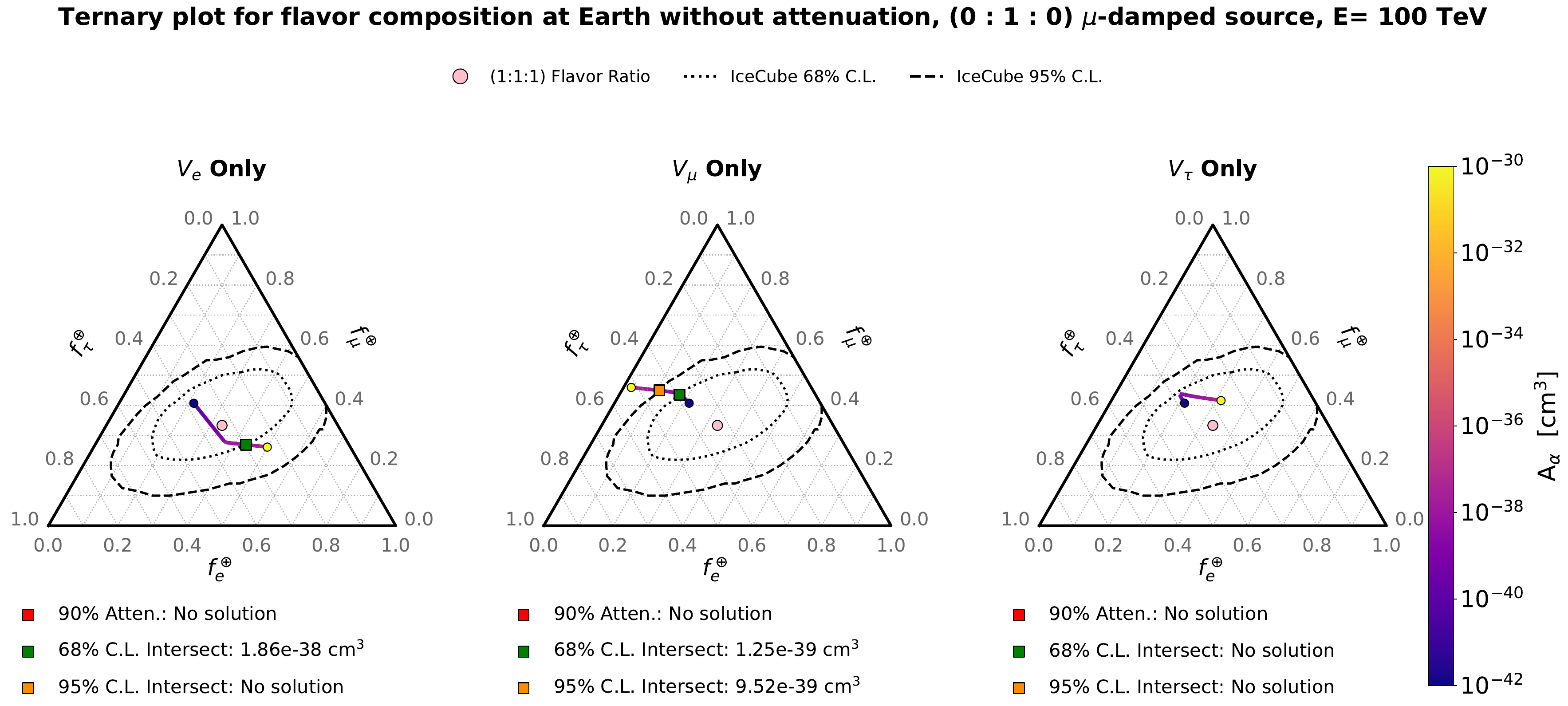}
    \caption{The same as Fig.~\ref{fig:120_no_attenuation} but for original $(0:1:0)$ flavor from $\mu$-damped source.
    }
    \label{fig:010_no_attenuation}
\end{figure}

We first switch off attenuation and ask how the potential alone changes the composition at Earth. We take the standard $p\gamma$ source, in which $p\gamma\to\Delta^+\to n\pi^+$ followed by $\pi^+\to\mu^+\nu_\mu$, $\mu^+\to e^+\nu_e\bar\nu_\mu$ gives $(f_e,f_\mu,f_\tau)^S=(\tfrac13,\tfrac13,0)$ for neutrinos and $(0,\tfrac13,0)$ for antineutrinos. 

Fig.~\ref{fig:120_no_attenuation} illustrates the resulting ternary trajectories for a standard $p\gamma$ astrophysical source (initial flavor ratio $1:2:0$) at $E = 100 \text{ TeV}$ under Normal Ordering, scanning the potential coefficient $A$ from $10^{-42}$ to $10^{-30} \text{ cm}^3$. We systematically study six scenarios, considering both opposite sign ($V_{\bar{\nu}} = -V_\nu$) and same sign ($V_{\bar{\nu}} = +V_\nu$) potentials coupled exclusively to a single lepton flavor ($V_e$, $V_\mu$, or $V_\tau$). We find that the flavor ratio are largely insensitive to same sign potentials across all flavors, as well as to opposite potentials coupled to $e$ or $\tau$; in these five configurations, the composition remains localized near the standard $(1:1:1)$ expectation and stays within the IceCube 68\% boundary. On the other hand, the $V_\mu$-only of opposite sign introduces a noticeable modification to the effective mixing matrix. In this case, the increasing potential drives a noticeable migration of the flavor composition outside the IceCube 68\% C.L. at coupling of $A = 8.41 \times 10^{-39} \text{ cm}^3$. The figure map the cross section (indicated by the color scale), marking critical phenomenological thresholds: the specific cross sections where the total flux suffers $90\%$ attenuation (red squares), and the precise geometric points where the modified flavor trajectory exits the IceCube $68\%$ (green squares) and $95\%$ (orange squares) C.L. in the figures.

In Fig.~\ref{fig:010_no_attenuation}, we next consider the muon-damped source scenario, wherein high-energy muons suffer severe radiative cooling or absorption prior to decay. In this regime, the production chain terminates at $\pi^+ \to \mu^+ \nu_\mu$, yielding an initial flavor composition $(f_e, f_\mu, f_\tau)^S = (0, 1, 0)$ consisting purely of muon neutrinos. In this case, there is only one row of plots since there is no anti neutrino. In this scenario, the standard vacuum oscillation expectation sits off-center, positioned within near the boundary of the IceCube 68\% C.L. For a $V_\tau$-only potential, the resulting ternary trajectory exhibits minimal geometric deviation, remaining safely confined within the 68\% C.L. region without any intersections. Conversely, potentials coupled to the electron or muon sectors induce significant phenomenological shifts to the effective mixing matrix. For the $V_e$-only case, the modified Hamiltonian drives the flavor fractions across the 68\% C.L. boundary at a potential strength of $A = 1.86 \times 10^{-38} \text{ cm}^3$. The most extreme structural deviation occurs in the $V_\mu$-only channel, where the trajectory migrates toward a $\nu_\tau$ and $\nu_\mu$-dominated composition. In this regime, the flavor fractions intersect the 68\% C.L. contour at $A = 1.25 \times 10^{-39} \text{ cm}^3$ and subsequently traverse the 95\% C.L. limit at $A = 9.52 \times 10^{-39} \text{ cm}^3$.

\section{Flavor-dependent attenuation}
\label{sec:attenuation}

\subsection{Formalism}
\label{sec:att_formalism}

Previous studies~\cite{Choi:2019ixb,Ferrer:2022xlw} of flavor-universal $\nu$--DM scattering constrain $\sigm$ by requiring that the surviving flux $\Phi^{\rm obs}/\Phi^{\rm ini}=\exp[-(\sigm)X_\chi]$ not fall below $10\%$, which with \eq{eq:column} gives $\sigm\lesssim6\times10^{-29}\,{\rm cm^2/GeV}$ for our benchmark $\Rem=0.023$~pc.\footnote{Ref.~\cite{Ferrer:2022xlw} quotes $2.0\times10^{-29}\,{\rm cm^2/GeV}$ with the same profile and the same 90\% criterion. The difference is the emission radius: that number corresponds to the lower end, $\Rem=10^{-2}$~pc, of the range $10^{-2}$--$1$~pc scanned there, for which we find $X_\chi=1.06\times10^{29}\,{\rm GeV/cm^2}$ and $\sigm\le2.2\times10^{-29}\,{\rm cm^2/GeV}$. Since $X_\chi\propto\Rem^{-4/3}$, moving from $0.023$ to $0.01$~pc strengthens the limit by a factor $3.0$. All our limits scale in the same way with $\Rem$.} In general the scattering cross section is flavor dependent, $\sigma_{\nu_\alpha\chi}$, in the same way as the potential; in a given model both are fixed by the coupling of $\nu_\alpha$ to the mediator, and the single-flavor scenarios $V_e$, $V_\mu$, $V_\tau$ only correspond to $\sigma_{\nu_\alpha\chi}=\sigma_{\nu\chi}\,\delta_{\alpha\beta}$ for $\beta=e,\mu,\tau$. In the decoherent limit the evolution of the density matrix $\varrho$ along the path, $d\varrho/dr=-i[\mathcal H,\varrho]-\tfrac12\{\Gamma,\varrho\}$ with
\begin{equation}
    \Gamma(r)=\frac{\rho_\chi(r)}{m_\chi}\,{\rm diag}\big(\sigma_{\nu_e\chi},\sigma_{\nu_\mu\chi},\sigma_{\nu_\tau\chi}\big),
    \label{eq:Gamma}
\end{equation}
reduces in the eigenbasis of $\mathcal H$ to independent damping of the diagonal entries,
\begin{equation}
    F^\oplus_i = F_i\,\exp\!\left[-\sum_\alpha\frac{\sigma_{\nu_\alpha\chi}}{m_\chi}\,X_{\alpha i}\right],
    \qquad
    X_{\alpha i}=\int_{\Rem}^{\infty}\rho_\chi(r)\,\big|(U_m(r))_{\alpha i}\big|^2\,dr ,
    \label{eq:attenuation}
\end{equation}
where $X_{\alpha i}$ 
is the column density weighted by the flavor-$\alpha$ content of the $i$-th eigenstate along the path. The flavor-universal case $\sigma_{\nu_\alpha\chi}=\sigma_{\nu\chi}$ recovers the usual $\exp[-(\sigm)X_\chi]$ for all eigenstates, independently of the mixing; the single-flavor scenarios give $F^\oplus_i=F_i\exp[-(\sigm)X_{\beta i}]$, and all the flavor dependence enters through the mixing in the medium. For antineutrinos the same expressions hold with $\bar U_m$ and $\bar F_i$, and with $\sigma_{\bar\nu_\alpha\chi}$, which we take equal to $\sigma_{\nu_\alpha\chi}$.

We assume that a scattered neutrino is removed from the observed flux; for light mediators the scattering is forward and a fraction of the flux is regenerated at lower energies~\cite{Murase:2019xqj}, which would weaken the limits. With attenuation, the flavor fluxes at Earth of \eqs{eq:fearth}{eq:fearthbar} become
\begin{equation}
    f^\oplus_\beta=\sum_i\big|(U_m^\oplus)_{\beta i}\big|^2\,F^\oplus_i ,
    \qquad
    \bar f^\oplus_\beta=\sum_i\big|(\bar U_m^\oplus)_{\beta i}\big|^2\,\bar F^\oplus_i ,
    \label{eq:fearth_att}
\end{equation}
where $\bar F^\oplus_i$ is given by \eq{eq:attenuation} with $\bar F_i$ and $\bar U_m$. The observable flux is  the sum $\Phi^\oplus_\beta=f^\oplus_\beta+\bar f^\oplus_\beta$ and its normalized fractions $f^{\rm obs}_\beta$ of \eq{eq:sum}, which are compared with the IceCube measurement; the total value $\sum_\beta\Phi^\oplus_\beta$ relative to its value without attenuation is used for the 90\% attenuation criterion.

\subsection{Eigenstate isolation at large potential}
\label{sec:isolation}

\begin{figure}[!t]
    \centering
    \includegraphics[width=1\linewidth]{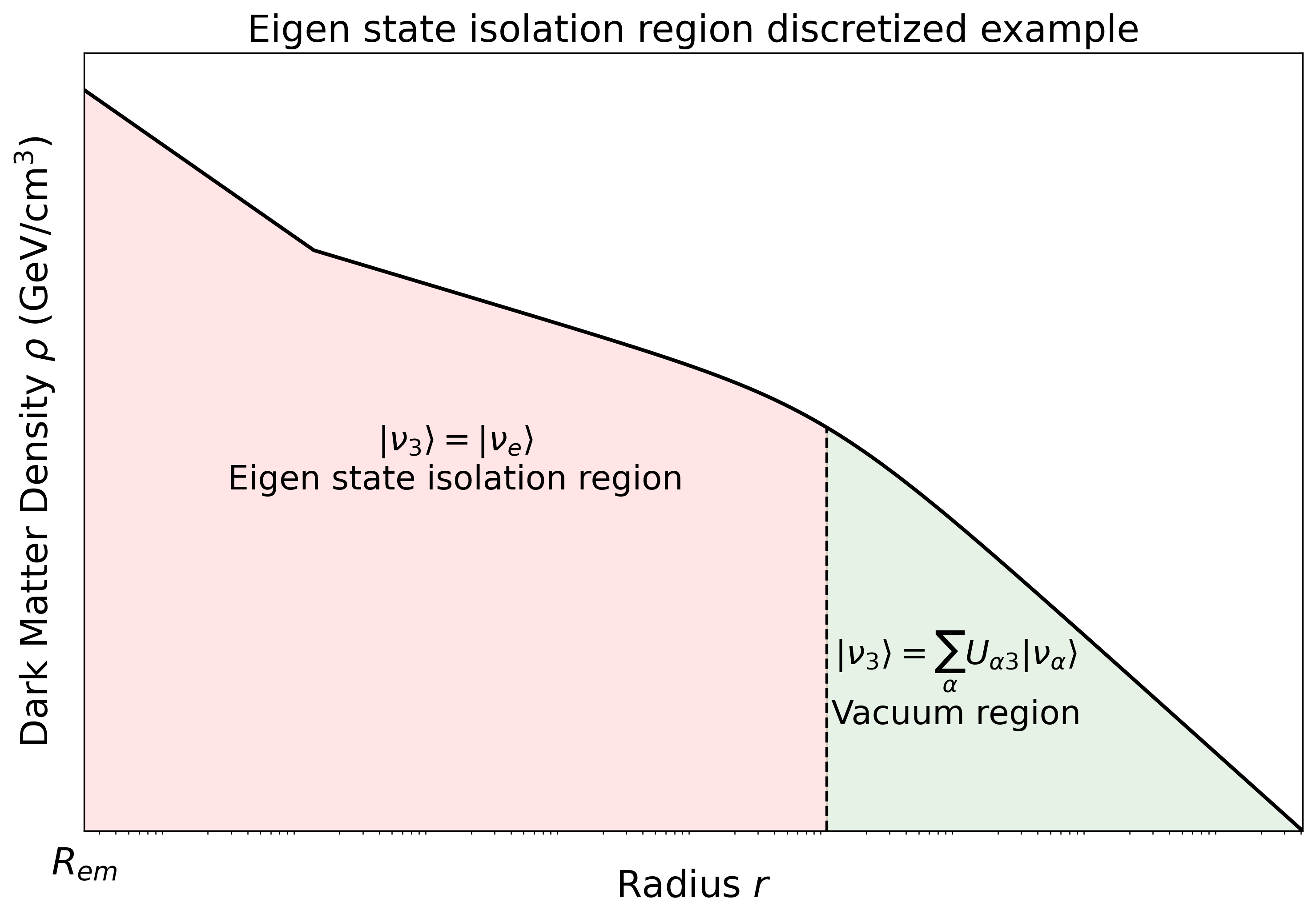}
    \caption{Illustration of the discretized two-zone approximation for eigenstate isolation along the dark matter density profile $\rho_\chi(r)$ from the emission radius $R_{\mathrm{em}}$ outward, shown for the $V_e$-only scenario ($V_e > 0$). In the inner high-density spike where $2EV_e(r) \gg \Delta m_{31}^2$ (pink shaded ``eigenstate isolation region''). In the outer halo where the dark matter density drops such that $2EV_e(r) \ll \Delta m_{21}^2$ (green shaded ``vacuum region'').}
    \label{fig:eigenstate isolation}
\end{figure}

When $2EV_e\gg\Delta m^2_{31}$ which is equivalent to $A\gg 10^{-36}\text{cm}^3$ for energy at 100TeV, the structure of $U_m^S$ can be read off analytically. With $V={\rm diag}(V_e,0,0)$ we have $U_{23}^\dagger VU_{23}=V$, so
\begin{equation}
    \mathcal H = U_{23}\Big[\tfrac{1}{2E}U_{13}U_{12}\mathcal M^2U_{12}^\dagger U_{13}^\dagger + V\Big]U_{23}^\dagger ,
\end{equation}
and $\theta_{23}$ is unaffected, $\theta^m_{23}=\theta_{23}$. In the 1--3 and 1--2 sectors the usual two-flavor formula $\tan2\theta^m=\Delta m^2\sin2\theta/(\Delta m^2\cos2\theta-2EV_e)$ gives $\theta^m_{13},\theta^m_{12}\to\pi/2$ for $V_e>0$ (neutrinos, or antineutrinos for same sign potentials). The mixing matrix at the source can be approximated as
\begin{equation}
    U_m^S \simeq
    \begin{pmatrix}
        0 & 0 & 1 \\
        -\cos\theta_{23} & -\sin\theta_{23} & 0 \\
        \sin\theta_{23} & -\cos\theta_{23} & 0
    \end{pmatrix}
    \label{eq:Um_large}
\end{equation}
and shows that the electron flavor is isolated in $\nu_3$, while the $\mu$ and $\tau$ flavors occupy the other two mass eigenstates.
In this approximation, two limits are instructive. For $A\ll10^{-38}\,{\rm cm^3}$, $U_m=U$ everywhere. Every mass eigenstate contains the interacting flavor and all three are attenuated. For $A\gg10^{-36}\,{\rm cm^3}$ the interacting flavor is isolated in a single mass eigenstate inside the spike. In this region only $\nu_3 \approx \nu_e$ is attenuated, while the other two are unattenuated. In the outer halo, where $\rho_\chi$ is many orders of magnitude smaller, $U_m$ becomes similar to $U$ and every mass eigen state can have attenuation. 

This two-zone behavior is illustrated in Fig.~\ref{fig:eigenstate isolation} and explains why the upper bounds on $\sigma_{\nu\chi}/m_\chi$ are raised at large potential. In the case of $A_\alpha \ll 10^{-38}\,\mathrm{cm}^3$, the matter mixing matrix equals the vacuum PMNS matrix ($U_m(r) \simeq U$) throughout the entire propagation path. Because every mass eigenstate contains a non-negligible admixture of the interacting flavor $\nu_\alpha$, all three eigenstates experience attenuation in the spike region, and are depleted as $\sigma_{\nu\chi}/m_\chi$ increases. Conversely, in the case of $A_\alpha \gg 10^{-36}\,\mathrm{cm}^3$, where $2EV_\alpha(r) \gg \Delta m_{31}^2$ in the inner spike, the mixing angles isolate the interacting flavor entirely into a single matter eigenstate, such as $|\nu_3\rangle \simeq |\nu_e\rangle$ in Eq.~\ref{eq:Um_large}, shown as the pink shaded region in Fig.~\ref{fig:eigenstate isolation}. Therefore, scattering within the spike attenuates only the single isolated eigenstate $\nu_3$, whereas the remaining two eigenstates $\nu_1$ and $\nu_2$ contain no interacting flavor component and escape the spike completely unattenuated. Because two of the three mass eigenstates survive the spike regardless of how large the cross section becomes, depleting $\nu_3$ alone is insufficient to reduce the total neutrino flux by $90\%$. In the outer vacuum region (the green shaded region in Fig.~\ref{fig:eigenstate isolation}), where $\rho_\chi(r)$ is small enough that $U_m(r) \to U$ and all three mass eigenstates once again share the interacting flavor. To significantly attenuate the surviving eigenstates $\nu_1$ and $\nu_2$, the cross section $\sigma_{\nu\chi}/m_\chi$ must be raised by several orders of magnitude until scattering becomes effective. 

\subsection{Attenuation curves and flavor trajectories for individual scenarios}
\label{sec:scenarios}

\begin{figure}[!t]
    \centering
    \includegraphics[width=1\linewidth]{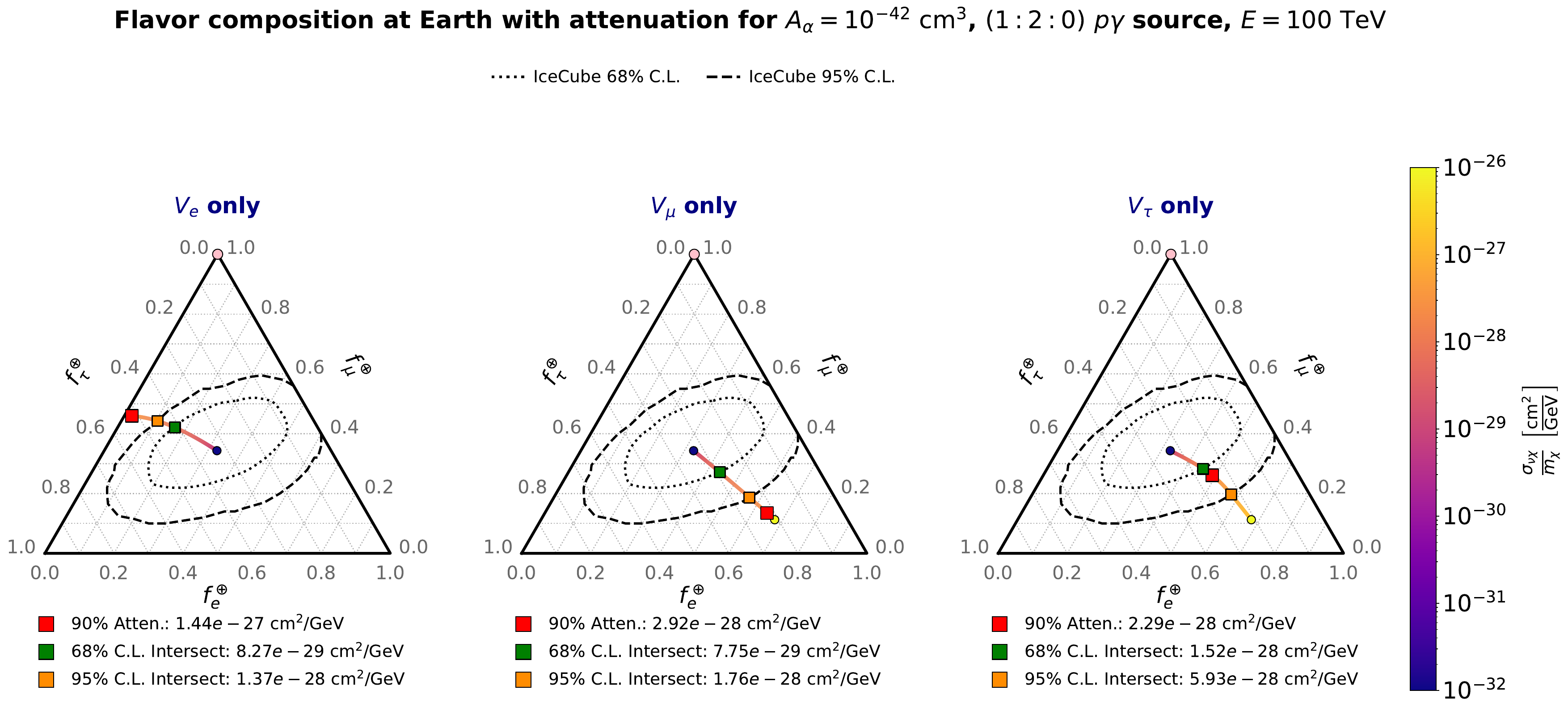}
    \caption{Flavor composition at Earth as a function of the scattering cross section $\sigma_{\nu\chi}/m_\chi$ (indicated by the color scale) in the vacuum-like regime ($A_\alpha = 10^{-42}~\mathrm{cm^3}$) for an initial $(1:2:0)$ $p\gamma$ source at $E = 100~\mathrm{TeV}$. The markers denote the specific cross-section values where the total flux is attenuated by 90\% (red squares) and where the flavor trajectory exits the IceCube 68\% (green squares) and 95\% (orange squares) C.L.\ contours. The panels show the single-flavor interaction scenarios $V_e$-only, $V_\mu$-only, and $V_\tau$-only from left to right. In this regime, the matter potential is negligible, so the flavor evolution is solely due to the scattering, resulting in the identical results for both opposite and same sign potentials.}
    \label{fig:120_attenuation_vacuum}
\end{figure}

\begin{figure}[!t]
    \centering
    \includegraphics[width=1\linewidth]{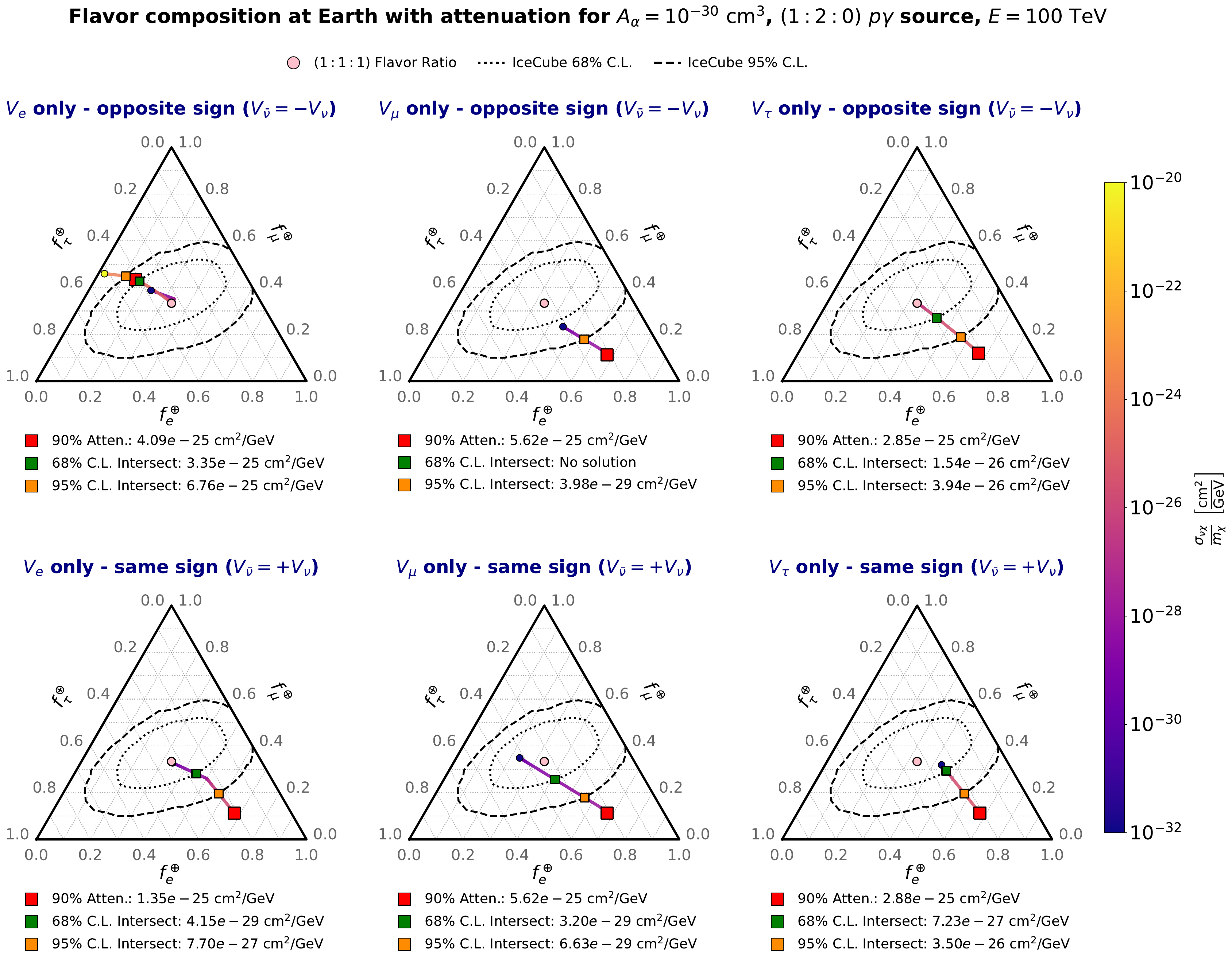}
    \caption{Trajectories of the observed flavor composition at Earth as a function of the scattering cross section $\sigma_{\nu\chi}/m_\chi$ (indicated by the color scale) in the eigenstate-isolation regime ($A_\alpha = 10^{-30}~\mathrm{cm^3}$) for an initial $(1:2:0)$ $p\gamma$ source at $E = 100~\mathrm{TeV}$. The markers denote the specific cross-section values where the total flux is attenuated by 90\% (red squares) and where the flavor trajectory exits the IceCube 68\% (green squares) and 95\% (orange squares) C.L.\ contours. The panels show the single-flavor interaction scenarios $V_e$-only, $V_\mu$-only, and $V_\tau$-only from left to right. The top row corresponds to opposite sign potential ($V_{\bar{\nu}} = -V_\nu$), while the bottom row corresponds to same sign potential ($V_{\bar{\nu}} = +V_\nu$).}
    \label{fig:120_attenuation_high}
\end{figure}

In this subsection we show, for each single-flavor interaction scenario, the flavor ratio at Earth as a function of $\sigm$ on the ternary diagram with the IceCube 68\% and 95\% CL. 

We extend our analysis by including scattering-induced attenuation for the standard $(1:2:0)$ $p\gamma$ source at $E = 100\,\mathrm{TeV}$, tracking the observed flavor ratio at Earth as a function of the scattering cross section $\sigma_{\nu\chi}/m_\chi$. Fig.~\ref{fig:120_attenuation_vacuum} illustrates the vacuum-like regime ($A_\alpha = 10^{-42}\,\mathrm{cm}^3$), where the matter potential is negligible everywhere ($U_m(r) \simeq U$) yielding identical plots for opposite and same sign potentials. In this limit, each mass eigenstate $i$ is attenuated independently. As $\sigma_{\nu\chi}/m_\chi$ increases, the eigenstates with the largest admixture of the interacting flavor $\alpha$ are depleted first. This explains why the trajectory of the $V_e$-only scenario moves in the opposite direction compared to the $V_\mu$-only and $V_\tau$-only scenarios. In the $V_e$-only case (left panel), because $|U_{e1}|^2 \simeq 0.68$ and $|U_{e2}|^2 \simeq 0.30$ are much larger than $|U_{e3}|^2 = \sin^2\theta_{13} \simeq 0.022$, scattering rapidly depletes $\nu_1$ and $\nu_2$ while leaving $\nu_3$ (and $\bar{\nu}_3$) comparatively unattenuated. Consequently, the surviving flux at Earth is dominated by $\nu_3$, shifting the flavor ratio leftward toward $(|U_{e3}|^2 : |U_{\mu 3}|^2 : |U_{\tau 3}|^2) \simeq (0.02 : 0.46 : 0.52)$, exiting the IceCube $68\%$ and $95\%$ C.L.\ at $\sigma_{\nu\chi}/m_\chi = 8.27 \times 10^{-29}\,\mathrm{cm^2/GeV}$ and $1.37 \times 10^{-28}\,\mathrm{cm^2/GeV}$, respectively, and reaching $90\%$ total flux attenuation at $1.44 \times 10^{-27}\,\mathrm{cm^2/GeV}$. Conversely, in both the $V_\mu$-only (middle panel) and $V_\tau$-only (right panel) scenarios, $\nu_1$ carries the smallest muon and tau fractions ($|U_{\mu 1}|^2 \simeq 0.11$ and $|U_{\tau 1}|^2 \simeq 0.21$), depleting $\nu_2$ and $\nu_3$ therefore leaves $\nu_1$ (and $\bar{\nu}_1$) as the dominant surviving state in both channels, driving the trajectories toward the bottom-right corner at $(|U_{e1}|^2 : |U_{\mu 1}|^2 : |U_{\tau 1}|^2) \simeq (0.68 : 0.11 : 0.21)$.

Fig.~\ref{fig:120_attenuation_high} depicts the eigenstate-isolation regime (refer to Sec.~\ref{sec:isolation}) , where $A_\alpha = 10^{-30}\,\mathrm{cm}^3$, for the $(1:2:0)$ $p\gamma$ source at $E = 100\,\mathrm{TeV}$. In contrast to the vacuum-like case (Fig.~\ref{fig:120_attenuation_vacuum}), the strong matter potential inside the spike isolates the interacting flavor into a single matter eigenstate, relaxing the $90\%$ flux attenuation bounds by roughly three orders of magnitude to $\sigma_{\nu\chi}/m_\chi \sim (1.35\text{--}5.62)\times 10^{-25}\,\mathrm{cm^2/GeV}$ and introduces a difference between opposite and same sign potentials in the $V_e$-only channel (left column). This difference originates from the opposite sign of the antineutrino potential ($V_{\bar{\nu}} = \mp V_\nu$) combined with the fact that the $p\gamma$ source produces neutrino flavor as neutrinos  $(f_e, f_\mu, f_\tau)^S = (1/3, 1/3, 0)$, and antineutrinos as $(\bar{f}_e, \bar{f}_\mu, \bar{f}_\tau)^S = (0, 1/3, 0)$. For neutrinos ($V_e > 0$), $\nu_e$ is isolated in $\nu_3^m$ above the $1\text{--}3$ resonance ($2EV_e \gg \Delta m_{31}^2$) and transfers into $\nu_2^m$ between the $1\text{--}3$ and $1\text{--}2$ resonances ($\Delta m_{21}^2 \ll 2EV_e \ll \Delta m_{31}^2$. In the same sign potential $V_e$-only scenario ($V_{\bar{\nu}} = +V_\nu$, bottom-left panel), antineutrinos experience the same positive potential, so $|\bar{\nu}_3^m\rangle \simeq |\bar{\nu}_e\rangle$ at the source; since $\bar{f}_e^S = 0$, $\bar{\nu}_3$ is unpopulated ($\bar{F}_3 = 0$) while $\bar{\nu}_1$ and $\bar{\nu}_2$ carry the $\bar{\nu}_\mu$ and  $\bar{\nu}_\tau$ flux. As $\sigma_{\nu\chi}/m_\chi$ increases, $\nu_3$ is depleted first in the spike (removing its vacuum projection $|U_{\beta 3}|^2$ at Earth, which is rich in $\nu_\mu$ and $\nu_\tau$), followed by the depletion of $\nu_2$ and $\bar{\nu}_2$ in the intermediate region, leaving $\nu_1$ and $\bar{\nu}_1$ as the surviving eigenstates. Consequently, the same sign potential $V_e$-only trajectory moves steadily toward the $\nu_1$ vacuum composition $(|U_{e1}|^2 : |U_{\mu 1}|^2 : |U_{\tau 1}|^2) \simeq (0.68 : 0.11 : 0.21)$ in the bottom-right corner, intersecting the $68\%$ C.L., $95\%$ C.L., and $90\%$ attenuation thresholds at $4.15 \times 10^{-29}$, $7.70 \times 10^{-27}$, and $1.35 \times 10^{-25}\,\mathrm{cm^2/GeV}$, respectively. On the other hand, in the opposite sign potential $V_e$-only scenario ($V_{\bar{\nu}} = -V_\nu$, top-left panel), the negative potential isolates $\bar{\nu}_e$ into the lightest matter eigenstate at the source, $|\bar{\nu}_1^m\rangle \simeq |\bar{\nu}_e\rangle$. Because $\bar{f}_e^S = 0$, $\bar{F}_1 = 0$ and the initial $\bar{\nu}_\mu$ flux instead populates $\bar{\nu}_2$ and $\bar{\nu}_3$. Thus, after $\nu_3$ is depleted in the spike and $\nu_2$, $\nu_1$, and $\bar{\nu}_2$ are attenuated in the outer halo, $\bar{\nu}_3$ remains as the longest-surviving state, driving the opposite sign $V_e$-only trajectory to the upper-left toward $(|U_{e3}|^2 : |U_{\mu 3}|^2 : |U_{\tau 3}|^2) \simeq (0.02 : 0.46 : 0.52)$, crossing the $68\%$ C.L.\ at $3.35 \times 10^{-25}\,\mathrm{cm^2/GeV}$, $90\%$ attenuation at $4.09 \times 10^{-25}\,\mathrm{cm^2/GeV}$, and the $95\%$ C.L.\ at $6.76 \times 10^{-25}\,\mathrm{cm^2/GeV}$. In the $V_\mu$-only (middle column) and $V_\tau$-only (right column) scenarios, $\nu_1$ and $\bar{\nu}_1$ remain the least-attenuated states in the outer halo, driving all trajectories toward the bottom-right with $90\%$ attenuation occurring at $5.62 \times 10^{-25}\,\mathrm{cm^2/GeV}$ for $V_\mu$-only and $(2.85\text{--}2.88) \times 10^{-25}\,\mathrm{cm^2/GeV}$ for $V_\tau$-only.

\begin{figure}[!t]
    \centering
    \includegraphics[width=1\linewidth]{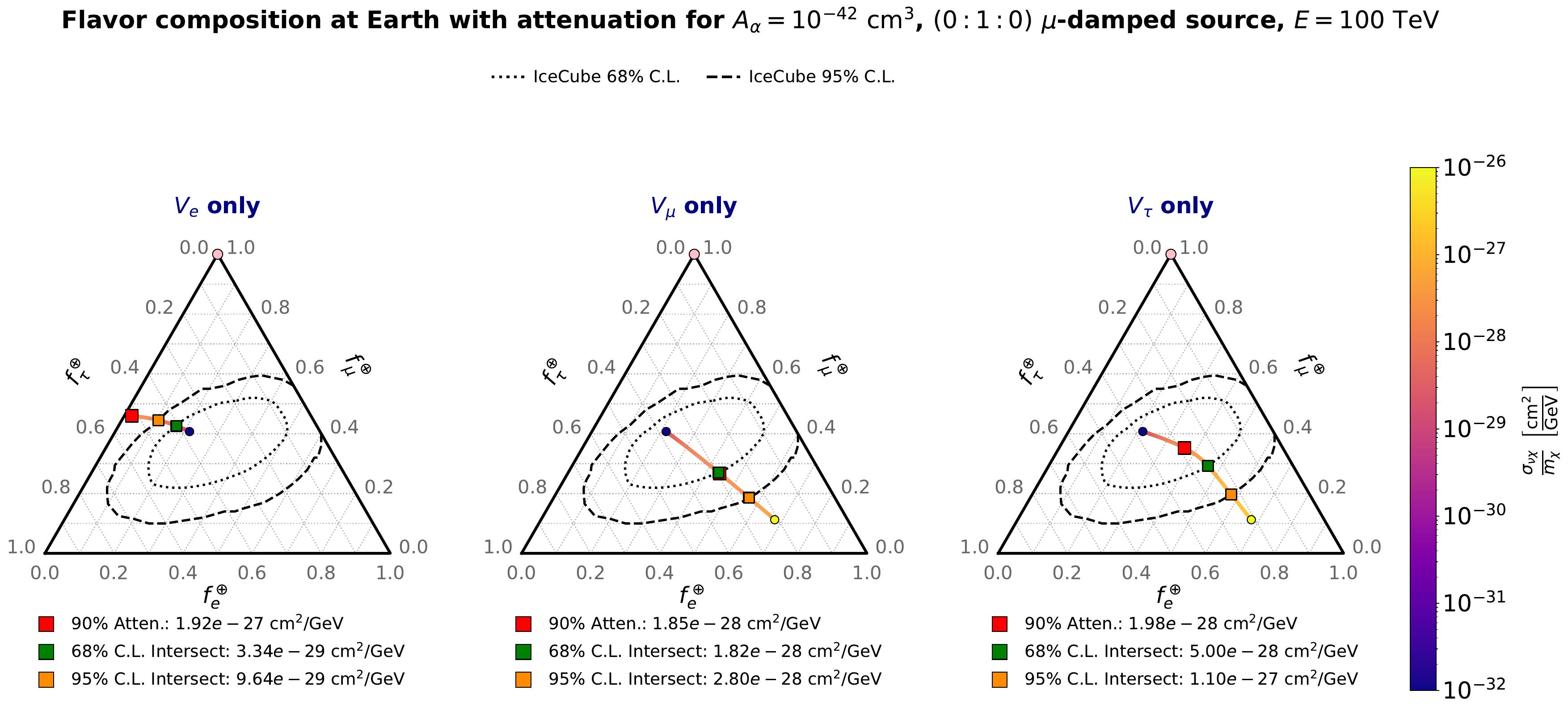}
    \caption{The same as Fig.~\ref{fig:120_attenuation_vacuum} but for an initial $(0:1:0)$ $\mu$-damped source. }
    \label{fig:010_attenuation_vacuum}
\end{figure}

\begin{figure}[!t]
    \centering
    \includegraphics[width=1\linewidth]{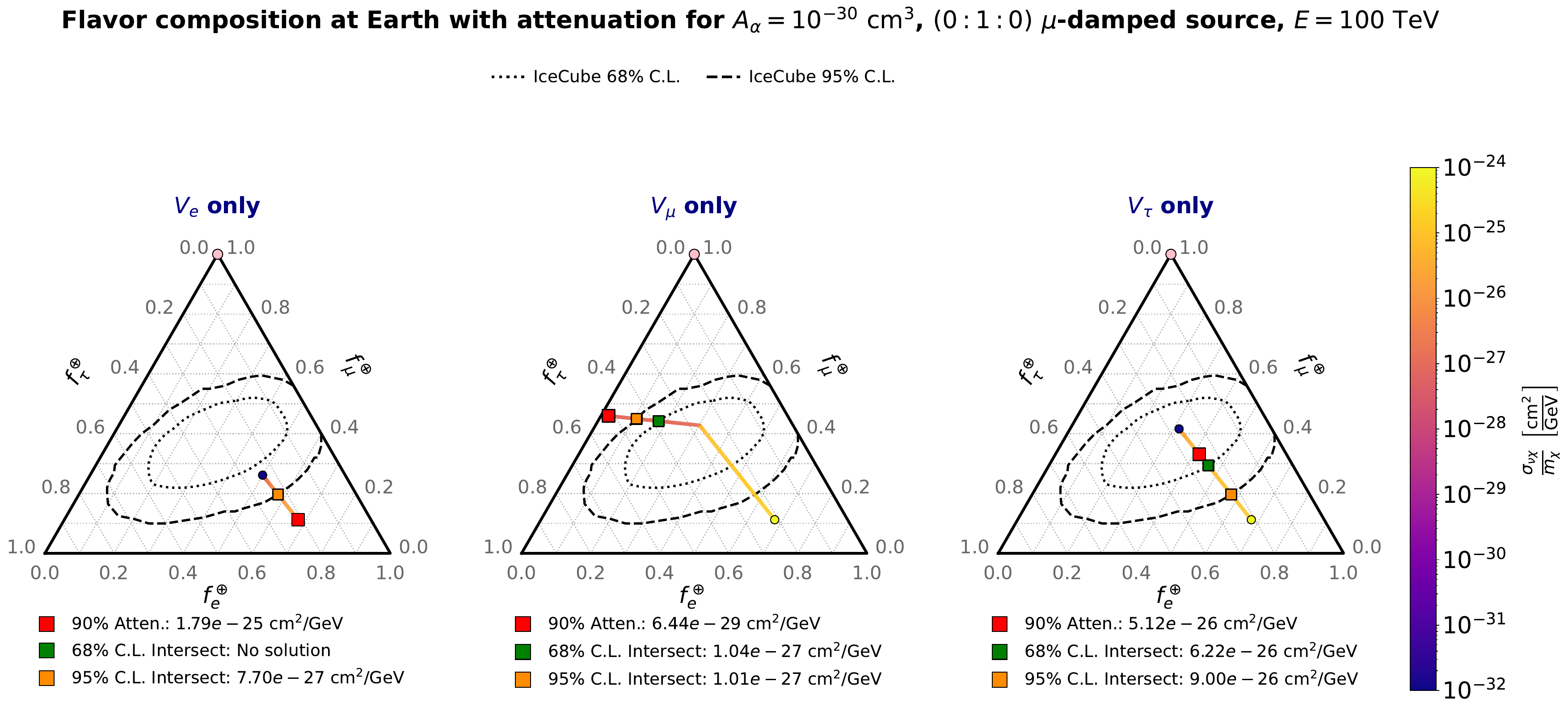}
    \caption{The same as Fig.~\ref{fig:120_attenuation_vacuum} but for an initial $(0:1:0)$ $\mu$-damped source.}

    \label{fig:010_attenuation_high}
\end{figure}

We further evaluate the impact of scattering-induced attenuation on a $\mu$-damped astrophysical source with an initial $(0:1:0)$ flavor ratio at $E = 100\,\mathrm{TeV}$, where the absence of antineutrinos renders the opposite and same sign potential scenarios identical. Fig~\ref{fig:010_attenuation_vacuum} shows the vacuum-like regime ($A_\alpha = 10^{-42}\,\mathrm{cm}^3$) and Fig.~\ref{fig:010_attenuation_high} shows the strongly matter-dominated eigenstate-isolation regime ($A_\alpha = 10^{-30}\,\mathrm{cm}^3$). Comparing the left panels of Figs.~\ref{fig:010_attenuation_vacuum} and~\ref{fig:010_attenuation_high} shows a reversal in the $V_e$-only trajectory: increasing $\sigma_{\nu\chi}/m_\chi$ drives the composition toward a low electron flavor fraction ($f_e^\oplus \to 0.02$) at $A_e = 10^{-42}\,\mathrm{cm}^3$, but toward a high electron flavor fraction ($f_e^\oplus \to 0.68$) at $A_e = 10^{-30}\,\mathrm{cm}^3$. At low potential ($A_e = 10^{-42}\,\mathrm{cm}^3$, Fig.~\ref{fig:010_attenuation_vacuum}), vacuum mixing at the source projects the initial $\nu_\mu$ flux across all three mass eigenstates with $F_i = |U_{\mu i}|^2 \simeq (0.14, 0.40, 0.46)$. Scattering therefore depletes $\nu_1$ and $\nu_2$ first and leaves $\nu_3$ as the surviving state, driving the Earth flavor ratio leftward to $|U_{\beta 3}|^2 \simeq (0.02 : 0.46 : 0.52)$ and crossing the $68\%$ C.L., $95\%$ C.L., and $90\%$ attenuation thresholds at $3.34 \times 10^{-29}$, $9.64 \times 10^{-29}$, and $1.92 \times 10^{-27}\,\mathrm{cm^2/GeV}$, respectively. Conversely, at high potential ($A_e = 10^{-30}\,\mathrm{cm}^3$, Fig.~\ref{fig:010_attenuation_high}), the dominant potential $2EV_e \gg \Delta m_{31}^2$ at the source isolates $\nu_e$ into $\nu_3^m$ according to Eq.~(4.5). Because the $(0:1:0)$ source contains no initial $\nu_e$, $\nu_3^m$ has zero initial population ($F_3 \simeq 0$), and the initial $\nu_\mu$ flux is shared strictly between $\nu_1$ and $\nu_2$. Both $\nu_1$ and $\nu_2$ escape the inner spike ($2EV_e \gg \Delta m_{31}^2$) unattenuated because $|(U_m)_{e1,2}|^2 \simeq 0$ in the spike. However, as the neutrinos propagate into the halo region between the atmospheric and solar resonances ($\Delta m_{21}^2 \ll 2EV_e(r) \ll \Delta m_{31}^2$), $\theta_{13}^m$ relaxes to $\theta_{13}$ while $\theta_{12}^m$ remains near $\pi/2$, so that $\nu_2^m$ becomes predominantly $\nu_e$ ($|(U_m)_{e2}|^2 \simeq \cos^2\theta_{13} \simeq 0.98$) whereas $\nu_1^m$ remains orthogonal to $\nu_e$ ($|(U_m)_{e1}|^2 \simeq 0$). This causes $\nu_2$ to be attenuated before $\nu_1$. The surviving flux at large cross sections is therefore dominated by $\nu_1$, which adiabatically connects to the lightest vacuum eigenstate at Earth and drives the flavor ratio toward the bottom-right corner. Finally, in the middle panel of Fig.~\ref{fig:010_attenuation_high} ($V_\mu$-only at $A_\mu = 10^{-30}\,\mathrm{cm}^3$), any potential exceeding $A_\mu = 9.52 \times 10^{-39}\,\mathrm{cm}^3$ already shifts the unattenuated flavor composition outside the IceCube $95\%$ C.L.\ contour toward the $\nu_\mu$--$\nu_\tau$ boundary (as seen in Fig.~\ref{fig:010_no_attenuation}), because the initial $\nu_\mu$ state is isolated into $\nu_3^m$ ($F_3 \simeq 1, F_{1,2} \simeq 0$) and arrives at Earth as $\nu_3$. Increasing $\sigma_{\nu\chi}/m_\chi$ attenuates this isolated $\nu_3$ component inside the spike (reaching $90\%$ flux attenuation already at $6.44 \times 10^{-29}\,\mathrm{cm^2/GeV}$). The C.L.\ intersections in this channel represent a minimum required cross section rather than an upper bound, although this regime is disfavored since more than $90\%$ of the total flux is already extinguished before the trajectory can re-enter the allowed IceCube region.


\begin{figure}[!t]
    \centering
    \includegraphics[width=1\linewidth]{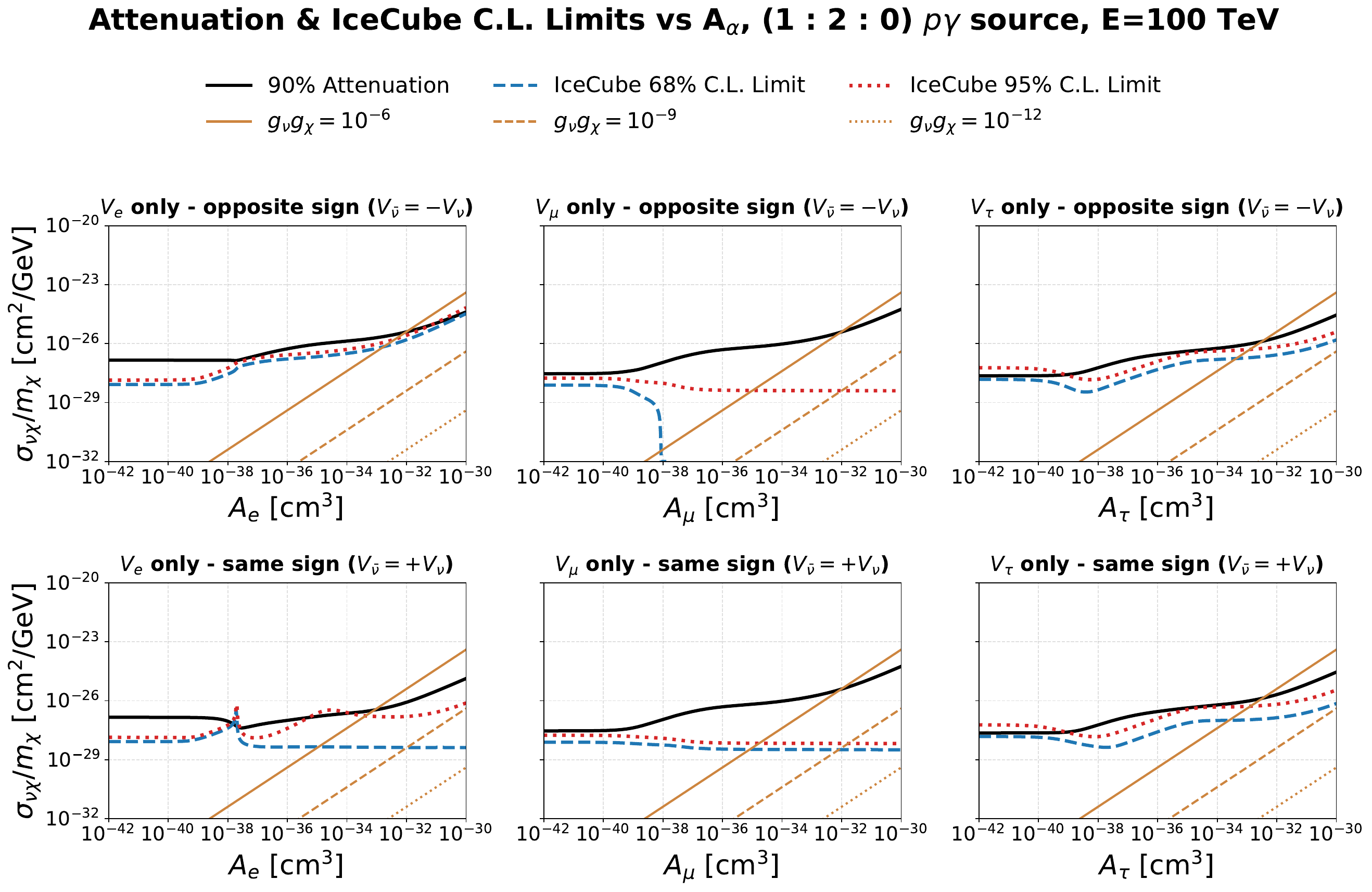}
    \caption{Upper limits on the dark matter--neutrino scattering cross section $\sigma_{\nu\chi}/m_\chi$ as a function of the effective potential parameter $A_\alpha$ for an initial $(1:2:0)$ $p\gamma$ flavor composition at $E = 100~\mathrm{TeV}$. The constraints are derived from the 90\% total flux attenuation criterion (solid black curves) and the IceCube 68\% (dashed blue curves) and 95\% (dotted red curves) C.L.\ flavor composition measurements. The panels display the single-flavor interaction scenarios $V_e$-only, $V_\mu$-only, and $V_\tau$-only from left to right. The top row assumes neutrinos and antineutrinos experience opposite potentials ($V_{\bar{\nu}} = -V_\nu$), while the bottom row assumes ($V_{\bar{\nu}} = +V_\nu$). The positively sloped orange lines represent the theoretical relation for a light mediator benchmark, $\sigma_{\nu\chi}/m_\chi \simeq (g_\nu g_\chi / 4\pi\hbar c)A_\alpha$, shown for coupling combinations $g_\nu g_\chi = 10^{-6}$, $10^{-9}$, and $10^{-12}$.}
    \label{fig:limits_A}
\end{figure}
\begin{figure}[!t]
    \centering
    \includegraphics[width=1\linewidth]{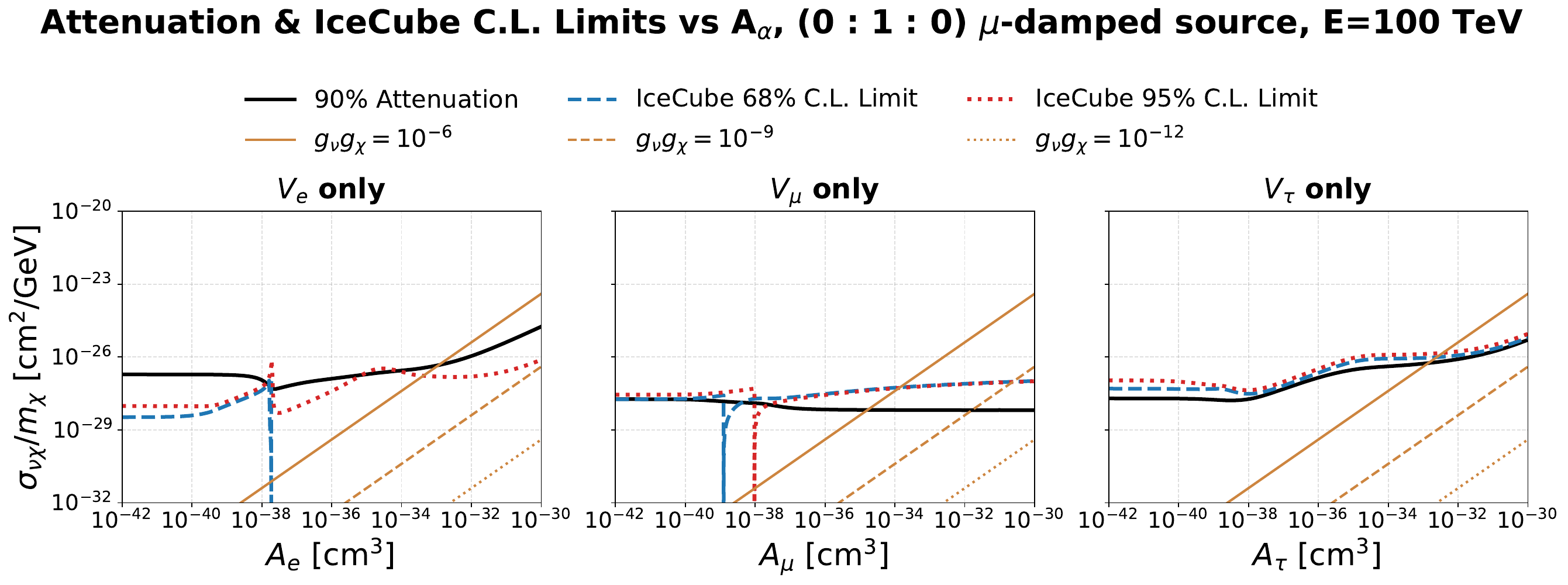}
    \caption{The same as Fig.\ref{fig:limits_A} but for an initial $(0:1:0)$ $\mu$-damped flavor composition. In the case of the middle panel ($V_\mu$ only), we have to say that for any value of $A_\mu$ bigger than $A_\mu = 9.52 \times 10^{-39}~\mathrm{cm}^3$, the 95\% C.L. limit curve actually represents the minimum value rather than the maximum.}

    \label{fig:limits_B}
\end{figure}

\subsection{Combined constraints and a mediator benchmark}
\label{sec:limits}

In Fig.~\ref{fig:limits_A}, we present the upper limits on the dark matter--neutrino scattering cross section $\sigma_{\nu\chi}/m_\chi$ as a function of the effective potential parameter $A_\alpha$ for an initial $(1:2:0)$ $p\gamma$ flavor composition at a neutrino energy of $E = 100\text{ TeV}$. The panels compare single-flavor interaction scenarios ($V_e$, $V_\mu$, and $V_\tau$ exclusively) for both opposite (upper panels) and same sign potentials (lower panels). For $A\ll 10^{-38}\text{cm}^3$, the bounds remain flat, dictating a strict cross-section limit of roughly $\sigma_{\nu\chi}/m_\chi \approx 10^{-27}\text{ cm}^2/\text{GeV}$ for $V_e$ only and $\sigma_{\nu\chi}/m_\chi \approx 10^{-28}\text{ cm}^2/\text{GeV}$ for $V_\mu$ or $V_\tau$ only, as shown in Sec.~\ref{sec:isolation}. However, as the potential parameter approaches the resonance region around $A_\alpha \sim 10^{-38}\text{ cm}^3$, matter effect begins to affect the bound on the cross section. For $A \gg 10^{-36}\text{ cm}^3$, where the eigenstate isolation begins to emerge in the spike, the 90\% total flux attenuation (solid black curves) rises with $A$, demonstrating that a much larger cross section is required to attenuate the flux when the interaction is dominated by the effective potential.

Fig.~\ref{fig:limits_B} shows the bound on the scattering cross section but for $\mu$-damped $(0:1:0)$  initial flavor composition. The 90\% attenuation bounds (solid black curves) remain qualitatively similar to those in Fig.~\ref{fig:limits_A} at small $A$. 
The most critical dynamical shift occurs in the middle panel representing the $V_\mu$-only interaction scenario. Specifically, for any potential value exceeding $A_\mu = 9.52 \times 10^{-39}\text{ cm}^3$, the 95\% C.L.\ limit curve transitions from representing a maximum allowed cross section to establishing a minimum required value. This implies that if the dark matter medium induces such a strong $V_\mu$ potential, the scattering cross section must be sufficiently large to attenuate the heavily distorted flux; otherwise, the predicted flavor ratios would violate IceCube observations.

The potential and the cross section are not independent in a given model. For a vector mediator exchanged in the $t$-channel, in the regime $2Em_\chi\ll m_{Z'}^2$ the total cross section is dominated by small momentum transfer and $\sigma_{\nu\chi}\simeq g_\nu^2g_\chi^2/(4\pi m_{Z'}^2)$, so that 
\begin{equation}
    \frac{\sigma_{\nu\chi}}{m_\chi}\simeq\frac{g_\nu g_\chi}{4\pi}\,A .
    \label{eq:model_line}
\end{equation}
In Fig.~\ref{fig:limits_A} and~\ref{fig:limits_B}, we show the excluded region in the $(A_e,\sigm)$ plane for the $V_e,V_\mu$ and $V_\tau$-only scenarios together with the lines of \eq{eq:model_line} for $g_\nu g_\chi=10^{-6},10^{-9},10^{-12}$ (orange solid, dashed and dotted lines respectively). The flavor-composition limit excludes the region above the solid line. For couplings $g_\nu g_\chi\lesssim10^{-12}$ the model lies below the attenuation bound for all $A$ allowed by \eq{eq:alpha_earth}.

\section{Discussion}
\label{sec:conclusion}

In this study, we have investigated the interplay between MSW effect and single  flavor selective attenuation due to dark matter on high-energy astrophysical neutrinos. We focused on neutrinos emitted deep within the dense dark matter spike surrounding the supermassive black hole of the blazar TXS 0506+056. We formulated single-flavor coupling ($\alpha = e, \mu, \tau$) generates both an MSW matter potential $V_\alpha(r) = A_\alpha \rho_\chi(r)$ and a flavor-specific scattering cross section $\sigma_{\nu_\alpha \chi}$. Because the dark matter density gradient along the line of sight satisfies the adiabaticity condition $\gamma_{\mathrm{ad}} \ll 1$, neutrino propagation preserves the mass-eigenstate from the production site to Earth. Imposing the consistency requirement that the local Galactic dark matter density ($\rho_\oplus = 0.4\text{ GeV/cm}^3$) not perturb the vacuum PMNS mixing matrix at $E = 100\text{ TeV}$ restricts the effective potential parameter to $A_\alpha \lesssim 10^{-30}\text{ cm}^3$.

Our analysis demonstrates that the inclusion of the matter potential alters cosmic neutrino flavor phenomenology across two distinct regimes separated by the MSW resonance threshold $A_\alpha \sim 10^{-38}\text{ cm}^3\,(100\text{ TeV}/E)$. In the case of $p\gamma$ source of original flavor (1: 2: 0), even in the absence of attenuation ($\sigma_{\nu\chi}/m_\chi \to 0$), a matter potential ($A_\alpha \gtrsim 10^{-38}\text{ cm}^3$) modifies the source mixing matrix $U_m^S$ and moves away from the $(1:1:1)$ Earth flavor ratio, but mostly stays in the IceCube 68\% and 95\% C.L. Similarly in the case of $\mu$-damped source of original flavor (0: 1: 0), with the MSW effect, the flavor ratio at Earth mostly stays inside 68\% and 95\% IceCube C.L. 

When dark matter--neutrino scattering is turned on, the movement of flavor ratio on the ternary plot becomes more noticeable, with large cross section of interaction in most of the cases the flavor ratio goes outside the IceCube 68\% and 95\% C.L. boundaries. The transition from $A_\alpha \lesssim 10^{-40}\text{ cm}^3$ (the vacuum-like regime) to  $A_\alpha \gg 10^{-36}\text{ cm}^3$ (the eigenstate-isolation regime) reshapes the upper bounds on $\sigma_{\nu\chi}/m_\chi$. In the vacuum-like regime, the interacting flavor is distributed across all three mass eigenstates throughout the spike, so scattering depletes all eigenstates and IceCube flavor and flux measurements impose limits of $\sigma_{\nu\chi}/m_\chi \lesssim 10^{-29}\text{--}10^{-28}\text{ cm}^2/\text{GeV}$. Conversely, in the eigenstate-isolation regime ($A_\alpha \gg 10^{-36}\text{ cm}^3$), the strong matter potential confines the interacting flavor into a single matter eigenstate inside the spike while leaving the other two eigenstates unattenuated. This eigenstate isolation creates a survival plateau that relaxes the 90\% total flux attenuation bound by up to four orders of magnitude---reaching $\sigma_{\nu\chi}/m_\chi \sim (1.4\text{--}5.6) \times 10^{-25}\text{ cm}^2/\text{GeV}$ at $A_\alpha = 10^{-30}\text{ cm}^3$.

Combining the 90\% flux attenuation criterion and the IceCube 68\% and 95\% C.L. flavor contours in the $(A_\alpha, \sigma_{\nu\chi}/m_\chi)$ plane we show a model-independent test for specific dark sector interaction, such as a light $t$-channel vector boson where $\sigma_{\nu\chi}/m_\chi \simeq (g_\nu g_\chi / 4\pi) A_\alpha$. While our benchmark constraints assume a dark matter spike profile characteristic of TXS 0506+056, this formalism can be readily adapted to other candidate neutrino-emitting active galactic nuclei, such as the Seyfert galaxy NGC 1068, by incorporating the corresponding black hole mass, emission radius, and spike index. Future extensions of this work should incorporate continuous neutrino energy spectra, energy-redistribution and regeneration effects from forward scattering via light mediators, and a full likelihood analysis combining source-specific flux normalizations with upcoming high-statistics three-flavor measurements from IceCube-Gen2 and KM3NeT.

\acknowledgments
The authors acknowledge the financial support from National Research Foundation (NRF) grant funded by the Korea government (MEST) NRF-2022R1A2C1005050.

\clearpage
\bibliographystyle{JHEP}
\bibliography{references}

\end{document}